\documentclass[preprint,12pt]{elsarticle}

\usepackage{graphicx}

\usepackage{amssymb}

\journal{Physics Letters A}

\begin{document}

\begin{frontmatter}



\title{Quantum dynamics of a nano-rod under compression}


\author{Geoffrey M. Beck}

\address{School of Chemistry and Physics, University of KwaZulu-Natal, 
Pietermaritzburg Campus,
Private Bag X01 Scottsville, 3209 Pietermaritzburg, South Africa,
E-mail: geoff.m.beck@gmail.com}

\author{Alessandro Sergi (Corresponding author)}

\address{
School of Chemistry and Physics, University of KwaZulu-Natal, Pietermaritzburg Campus,
Private Bag X01 Scottsville, 3209 Pietermaritzburg, South Africa,
E-mail: sergi@ukzn.ac.za, Tel: +27 33 260 5875, Fax: +27 33 260 5876 }

\begin{abstract}
A nano-rod under compression, which 
is capacitively coupled to a Cooper-pair
box, can be modeled in terms of a quartic oscillator
linearly interacting with a tunnelling pseudo-spin.
We have integrated numerically
the quantum dynamics of this system 
in the partial Wigner representation and calculated
the pseudo-spin population difference.
Depending on the coupling, we have verified that the
quantum tunnelling of the oscillator can lead
to a more effective reduction of the
Rabi oscillation amplitude of the pseudo-spin.
Such findings suggests an experimental set-up that
might be able to discriminate between the quantum
and the classical motion of a nano-rod.
\end{abstract}


\end{frontmatter}

% \linenumbers

\section{Introduction}
\label{}

While there has been substantial progress in cooling mechanical
oscillators close to the point where quantum-mechanical
zero-point effects are 
important~\cite{Wilson-Rae,Marquardt,Rocheleau,OConnell,Teufel,Chan},
the problem of witnessing
the quantum mechanical features in an unambiguous way
remains somewhat unsolved.
Solution of such a problem is desirable since
the observation of quantum effects in mesoscopic mechanical 
human-made objects could lead to testing the basic principles
of quantum mechanics~\cite{Blencowe,Schwab,Adler}. 
With respect to this,
it has been known since the beginning of quantum mechanics
that distinguishing between quantum and classical features of
isolated harmonic oscillators is difficult.
For example, the law of evolution of a quantum harmonic
oscillator in the Wigner formulation of quantum mechanics~\cite{wigner}
looks perfectly classical.
Nevertheless, theoretical studies have been performed both on non-linear 
oscillators~\cite{Katz,Katz2,Claudon,Guo,Rips}
and harmonic oscillators~\cite{Armour,Irish,Rakesh,Brouard}.
However,
when a nano-rod is under longitudinal compression, the potential energy profile
can vary from a harmonic form to that of a double well~\cite{Chakraborty,Chakraborty2}.
The two minima in the wells describe the two 
buckled states, which can be obtained at a particular strain~\cite{Carr}.
In such a potential energy profile, the nano-rod can move from
one well minima to the other either by thermal fluctuations
or by quantum tunnelling. 
One can expect that, for such non-linear dynamics,
quantum effects might be less ambiguous to detect.
For example, in order to observe the quantum motion of
the nano-rod, one could couple it
capacitively to a Cooper-pair box~\cite{Makhlin}
(following what other authors
have considered in the case of harmonic 
nano-oscillators~\cite{Armour,Irish,Rakesh}).
A Cooper-pair box is composed of a small superconducting island
weakly linked to a superconducting reservoir~\cite{Makhlin}.
The balance between the charging energy and the tunneling strength
of the Cooper-pair between the island and the reservoir controls
the state of the Cooper-pair box.
External gates can drive the Cooper-pair box into either its ground or excited
state, with definite Cooper-pair numbers, or into a linear combination
of such states. Such systems are possible candidates as controllable qubits
in quantum computing devices~\cite{Makhlin}.
 
\section{Theory}

The system comprised by the non-linear oscillating nano-rod coupled
to the Cooper-pair box can be modelled by a
a quartic oscillator linearly
coupled to a tunnelling pseudo-spin.
In this Letter, we have integrated numerically the quantum dynamics
of the total system and looked for signatures of the quantum evolution
of the non-linear oscillator in the reduced density matrix of
the pseudo-spin. To this end, we adopted a mixed Wigner representation
of quantum mechanics~\cite{ilya} and described the quartic oscillator
in phase space. In such a representation, the quantum effects on the
evolution of the quartic oscillator can be suitably obtained
in terms of the higher order derivatives of the partially
Wigner represented density matrix (PWRDM) $\hat{W}(R,P,t)$.

%%%%%%%%%%%%%%%%%%%%%%%%%%%%%%%%%%%%%%%%%%%%%%%%%%%%%%%%%%%%%%%%%%%%

%%%%%%%%%%%%%%%%%%%%%%%%%%%%%%%%%%%%%%%%%
%
The PWRDM is defined as:
\begin{eqnarray}
\hat{W}(R,P,t)
&=&\frac{1}{2\pi\hbar}
\int dz e^{iP\cdot z/\hbar}
\langle R-\frac{z}{2}\vert\hat{\rho}
\vert R+\frac{z}{2}\rangle \;.
\nonumber \\
\end{eqnarray}
The transform of operators is defined analogously.
In such a representation, the law of motion is written as
\begin{eqnarray}
\frac{\partial}{\partial t}\hat{W}(R,P,t)
&=&-\frac{i}{\hbar}
\left[\hat{H}_W
\exp\left(\frac{i\hbar}{2}
\overleftarrow{\mbox{\boldmath$\partial$}}
\cdot\mbox{\boldmath$\epsilon$}\cdot
\overrightarrow{\mbox{\boldmath$\partial$}}
\right)\hat{W}
\right.\nonumber\\
&-&
\left.
\hat{W}
\exp\left(\frac{i\hbar}{2}
\overleftarrow{\mbox{\boldmath$\partial$}}
\cdot\mbox{\boldmath$\epsilon$}\cdot
\overrightarrow{\mbox{\boldmath$\partial$}}
\right)
\hat{H}_W
\right]\;.
\label{eq:pW}
\end{eqnarray}
In Eq.~(\ref{eq:pW})
$\mbox{\boldmath$\epsilon$}=-\mbox{\boldmath$\epsilon$}^T$
is the total antisymmetric symbol
in phase space, so that $\overleftarrow{\mbox{\boldmath$\partial$}}
\cdot\mbox{\boldmath$\epsilon$}\cdot
\overrightarrow{\mbox{\boldmath$\partial$}}$
denote the Poisson bracket operator~\cite{b3}.
The symbol
$\hat{H}_W$ is the partially Wigner transformed Hamiltonian,
which in this Letter is defined as
\begin{equation}
\hat{H}_W=-\Omega\hat{\sigma}_x-cR\hat{\sigma}_z+H_{C,W}\;.
\label{eq:tot-ham}
\end{equation}
where $\hat{\sigma}_x$ and $\hat{\sigma}_z$ are Pauli spin operators.

Quantum averages in the partial Wigner representation of quantum mechanics
are calculated as
\begin{equation}
\langle\hat{O}(t)\rangle
=
{\rm Tr}'\int dRdP \hat{O}_W(R,P)\hat{W}(R,P,t)\;.
\end{equation}
%%%%%%%%%%%%%%%%%%
%
In the general case, Eq.~(\ref{eq:pW}) is not simpler to solve
that the standard Von Neumann equation~\cite{petruccione}.
However, 
when the Hamiltonian $H_{C,W}$ has the form
\begin{equation}
H_{C,W}=
\frac{P^2}{2}+\sum_{k=1}^{n_k}\frac{b_k}{k!}R^k\;,
\label{eq:hambqua}
\end{equation}
where $n_k$ is a fixed integer, Eq.~(\ref{eq:pW}) becomes
\begin{eqnarray}
\frac{\partial}{\partial t}\hat{W}(R,P,t)
&=&-\frac{i}{\hbar}
\left[\hat{H}_W
\left( 1+{\cal A} \right)
\hat{W}
\right.\nonumber\\
&-&
\left.
\hat{W}
\left( 1+{\cal A} \right)
\hat{H}_W
\right]\;,
\label{eq:pW2}
\end{eqnarray}
where ${\cal A}$ is a differential operator of at most the same order
$n_k$ as the polynomial appearing in Eq.~(\ref{eq:hambqua}).
Its definition is given by
\begin{equation}
{\cal A} = \sum_{n=1,3,5, ...}^{n_k}\frac{1}{n!}
\left(\frac{i\hbar}{2}\right)^n
\left( \overleftarrow{\mbox{\boldmath$\partial$}} \cdot
\mbox{\boldmath$\epsilon$}\cdot
\overrightarrow{\mbox{\boldmath$\partial$}}\right)^n
\;.
\label{eq:defA}
\end{equation}
Here, we define a quartic oscillator by setting $n_k=4$
in Eq.~(\ref{eq:hambqua}).
Quantum dynamics is obtained upon considering all
the quantum corrections  in Eq.~(\ref{eq:defA}).
while the classical evolution of the oscillator
(coupled to the quantum pseudo-spin)
is obtained by considering only n=1 in Eq.~(\ref{eq:defA}).
This latter is a known
hybrid scheme of motion~\cite{qcl5,qcl6,qcl7,qcl8}
according to which
the quartic oscillator evolves classically on the two adiabatic potential 
energy surfaces determined by the Cooper-pair box system.

\section{Model and numerical calculations}

When represented in the basis of the eigenvectors
of $\Omega\hat{\sigma}_x$, Eq.~(\ref{eq:pW2}) appears
as a set of coupled linear partial differential equations
defining an initial value problem for the PWRDM.
We have studied a model defined by the Hamiltonian
in Eq.~(\ref{eq:tot-ham}) with Eq.~(\ref{eq:hambqua})
specialised to
\begin{equation}
H_{C,W}=\frac{P^2}{2}+\frac{b_2}{2}R^2+\frac{b_4}{4}R^4\;,
\label{eq:HCW}
\end{equation}
where all quantities in the Hamiltonian appear in dimensionless form. 
One can obtain the model Hamiltonian in Eq.~(\ref{eq:HCW})
following the analysis performed in~\cite{Carr}.
According to this, $H_{C,W}$ in Eq.~(\ref{eq:HCW}) 
describes the dynamics of the fundamental mode
of oscillation of the nano-rod:
the variable $R$ denotes the Fourier amplitude
of the fundamental displacement
and $P$ is the momentum conjugated to such a displacement.
The dimensionless variables appearing in Eq.~(\ref{eq:HCW})
are related to their dimensionful counterparts by:
\begin{eqnarray}
R &=& \left( \frac{M \omega_0}{\hbar}\right)^{1/2} R^{\prime} 
\label{eq:dim1}\; \\ 
P &=& (\hbar M \omega_0)^{-1/2} P^{\prime} \; ,
\label{eq:dimless-vars}
\end{eqnarray}
Additionally, dimensionless parameters are similarly defined:
\begin{eqnarray}
\Omega &=& \frac{\Omega^{\prime}}{\omega_0} \;, \\
c &=& \frac{c^{\prime}}{\omega_0\sqrt{M \omega_0 \hbar}} \;,  \\
b_2 &=& \frac{1}{M \omega_0^2} b_2^{\prime} \;, \\
b_4 &=& \frac{\hbar}{M^2 \omega_0^3} b_4^{\prime}
\label{eq:dimn} \;. 
\end{eqnarray}
Here, primed variables denote dimensionful quantities.
In this notation $\omega_0$ and $M$ are the fundamental frequency and mass of 
the quartic oscillator, entering the definition of both
$b_2^{\prime}$ and $b_4^{\prime}$, and are also dimensionful.
The dimensionless coefficient $b_2$ is related to the strain $\epsilon$ by~\cite{Carr} 
\begin{equation}
b_2=\frac{\epsilon_c-\epsilon}{\epsilon_c}\;,
\end{equation}
where $\epsilon_c$ is the critical strain, such that
when $\epsilon>\epsilon_c$ one has a double well
potential profile for the oscillator.
Setting the fundamental frequency $\omega_0 = 0.5$ GHz
and the mass $M = 10^{-21}$ kg, similar to those discussed
in~\cite{Carr} for a silicon nano-rod,
the time ($t=\omega_0t^{\prime}$)
is given in ns and the length scale of the nano-rod
is approximately $0.1$ \AA.
In the following, we will refer to the dimensionless quantities.
However, the use of Eqs.~(\ref{eq:dim1}-\ref{eq:dimn}) allows
one to easily determine the physical values.

In this study, we use two sets of values for the quantities
describing the elastic properties of the nano-rod.
The first set is given by $b_2=-1$ corresponding to
$b_2'=-10^{-5}$ kg s$^{-2}$ and $b_4=0.5$ corresponding
to $b_4'= 4.742 \times 10^{15}$ kg m$^{-2}$ s$^{-2}$.
This set of values has allowed us to observe relevant 
dynamical effect on a short time scale.
The second set chosen to be in agreement with the
prediction of the elastic theory of materials for a nano-rod
with the values of frequency and mass given above
 and thickness of the order of the nano-meters~\cite{Carr};
in this case $b_2=-0.01$, corresponding to $b_2'=-1\times 10^{-7}$
kg s$^{-2}$, and $b_4=0.0004$, corresponding to  $b_4'=4.742\times 10^{14}$
kg m$^{-2}$ s$^{-2}$.
This second set of values has required us to extend the calculation
in the time range for almost a factor of two.

For the numerical calculation we 
have employed the method of lines \cite{pdes} 
by discretizing phase space using a grid of 120 points
in both the $R$ and $P$ directions.
The resulting system of ordinary differential equations, in the time variable,
has been integrated using a Runge-Kutta Cash-Karp~\cite{cash-karp} integrator of fifth order,
with an time-step of $10^{-4}$ in dimensionless units.
The time scale achieved in our calculations (of the order of $10^1$ - $10^2$
ns) is satisfactory, 
in that the decoherence time of a suitably prepared Cooper-pair box
is of the order of $500$ ns~\cite{Irish}.
In our numerical simulations, we have considered an initial PWRDM
with the spin in its excited state and the oscillator in a coherent state.
The PWRDM is then explicitly given by
\begin{eqnarray}
\hat{W}(R,P) &=& \left[\begin{array}{cc} 1 & 0 \\ 0 & 0\end{array}\right]
\times \frac{1}{\pi\hbar}
\exp[-\frac{(R-R_0)^2}{2(\Delta R)^2}] \nonumber\\
&\times& \exp[-2(\Delta R)^2(P-P_0)^2 ]\;,
\end{eqnarray}
where $\Delta R $ represents the uncertainty in the position of the oscillator.
In our calculations, we have  chosen the particular
value of $\Delta R$ to be 0.6071, 
to aid initial confinement of the oscillator
to a single well of the potential profile.
Additionally, the parameter $R_0$ has been chosen so that the coherent state is initially
located in the ground state
minimum of the left well of the quartic potential.
The numerical simulation of Eq.~(\ref{eq:pW2})
can be performed with or without the inclusion of the
quantum corrections on the dynamics of the quartic oscillator, corresponding
to the full quantum and hybrid cases respectively.

We have performed a series of calculations as a function of the 
coupling parameter $c$ between the pseudo-spin and the quartic oscillator
($0.1 \le c \le 1$). In such a range of parameters we have examined
the effect of the quantum nature of the evolution of the
quartic oscillator on the pseudo-spin population difference $\langle\hat{\sigma}_z(t)\rangle$ 
and the position-probability Prob$(R)$ of the oscillator itself. 

For $b_2=-1$ and $b_4=0.5$,
in the time interval $t \leq 10$, we observed that the inclusion
of quantum dynamical effects promotes tunnelling between
the wells of the potential profile. 
In turn, this leads to modifications, based on the width 
of the oscillator distribution in phase space,
of the effective damping experienced by the Rabi oscillations 
in the population difference. 
These differences allow the quantum case to damp the Rabi oscillations more
effectively at short times but also allow for the possibility of promoting
Rabi oscillations if longer times are studied. 
These dual possibilities exist because 
the tunnelling introduced by the quantum dynamical effects 
allows the oscillator distribution to shift continuously between being 
highly symmetric (occupying both wells) and being asymmetric 
(largely occupying a single well).
Whereas the classical oscillator distribution was
seen to be largely static after short times.

\begin{figure}
\includegraphics[scale=0.7]{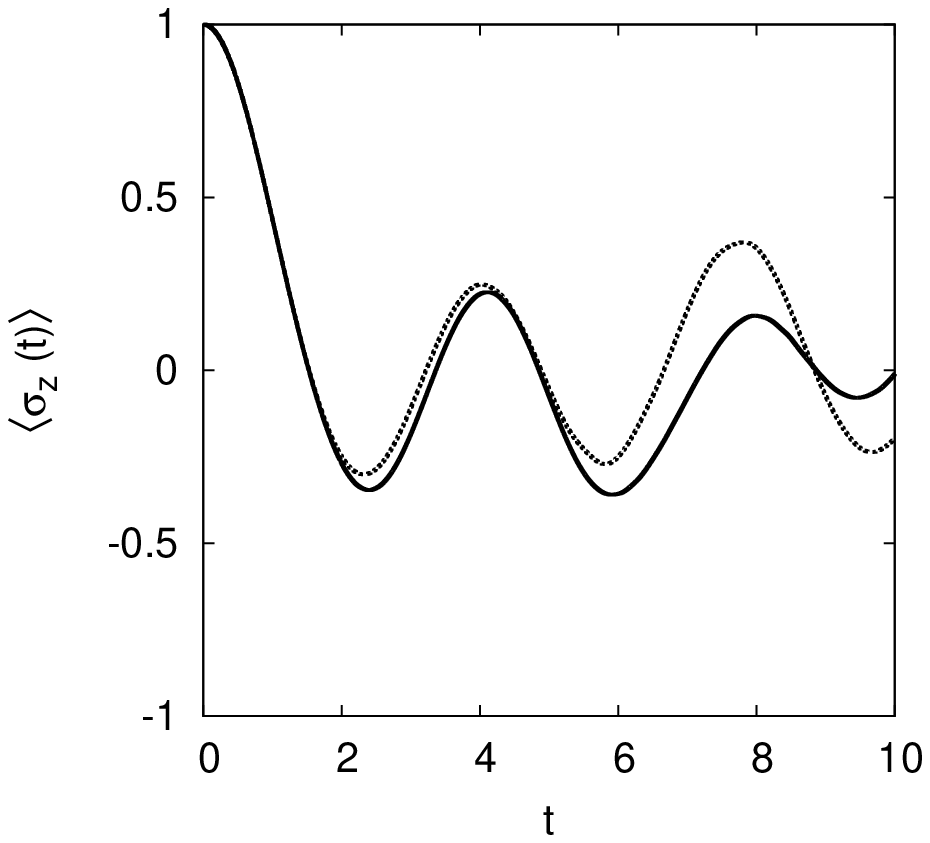}
\caption{Time dependence of the population $\langle \sigma_z(t) \rangle$. 
The dotted line corresponds to the quantum-classical evolution
while the solid-line corresponds to the quantum-corrected evolution.
Frequency ratio $\Omega = 0.6$, coupling constant $c = 0.4$,
initial average position $R_0 = -1.6$, harmonic coefficient $b_2 = -1$
and non-linearity parameter $b_4 = 0.5$.
}
\label{fig:fig1}
\end{figure}

\begin{figure}
\includegraphics[scale=0.7]{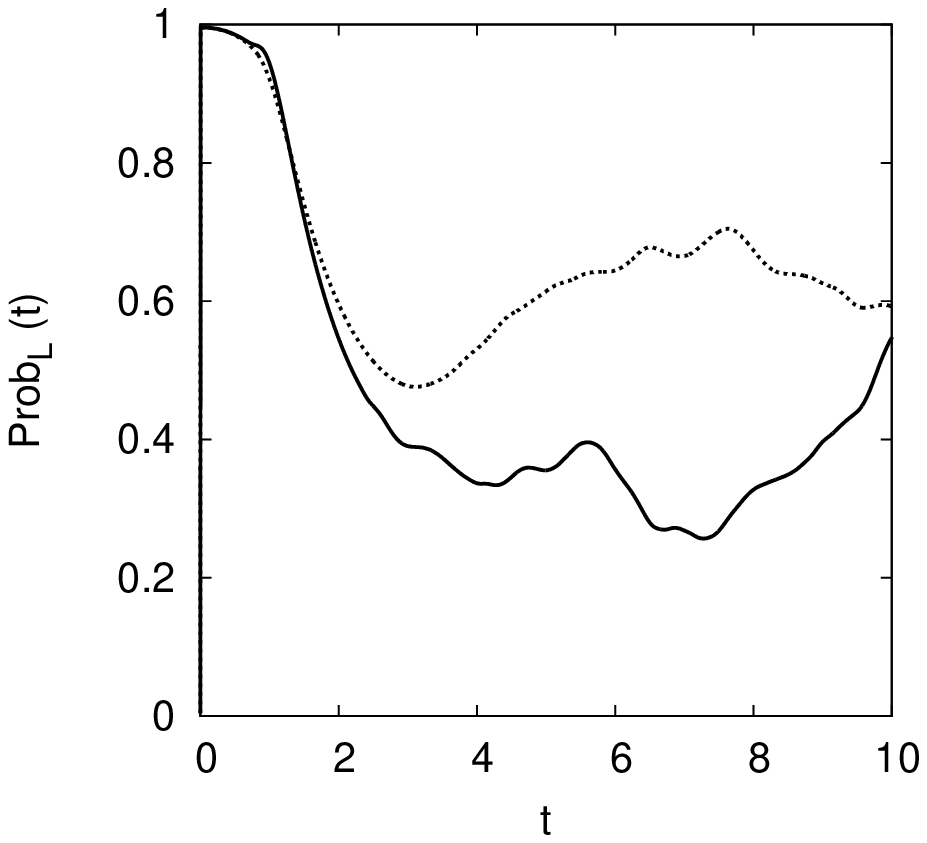}
\caption{Time dependence of instantaneous oscillator left-well 
occupation probability Prob$_L$($t$).
The dotted line corresponds to the quantum-classical evolution
while the solid-line corresponds to the quantum-corrected evolution.
Frequency ratio $\Omega = 0.6$, coupling constant $c = 0.4$,
initial average position $R_0 = -1.6$, harmonic coefficient $b_2 = -1$ 
and non-linearity parameter $b_4 = 0.5$}
\label{fig:fig2}
\end{figure}

\begin{figure}
\includegraphics[scale=0.7]{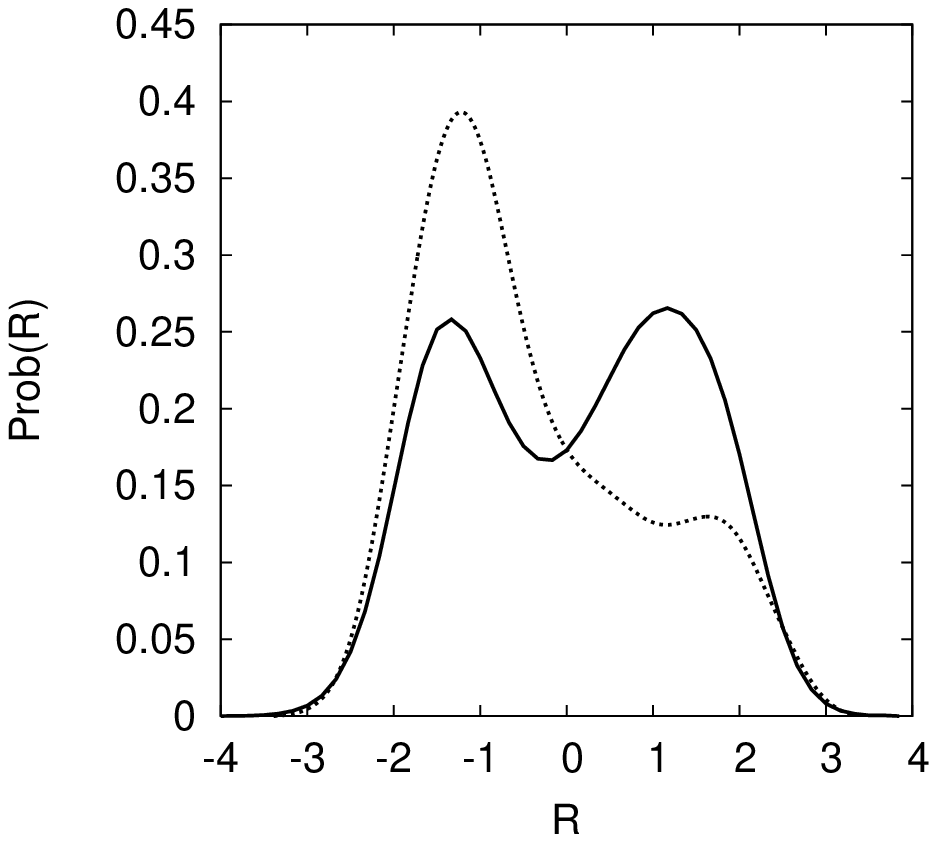}
\caption{Oscillator position-probability Prob($R$).
The dotted line corresponds to the quantum-classical evolution
while the solid-line corresponds to the quantum-corrected evolution.
Frequency ratio $\Omega = 0.6$, coupling constant $c = 0.4$,
initial average position $R_0 = -1.6$, harmonic coefficient $b_2 = -1$ 
and non-linearity parameter $b_4 = 0.5$}
\label{fig:fig3}
\end{figure}

\begin{figure}
\includegraphics[scale=0.7]{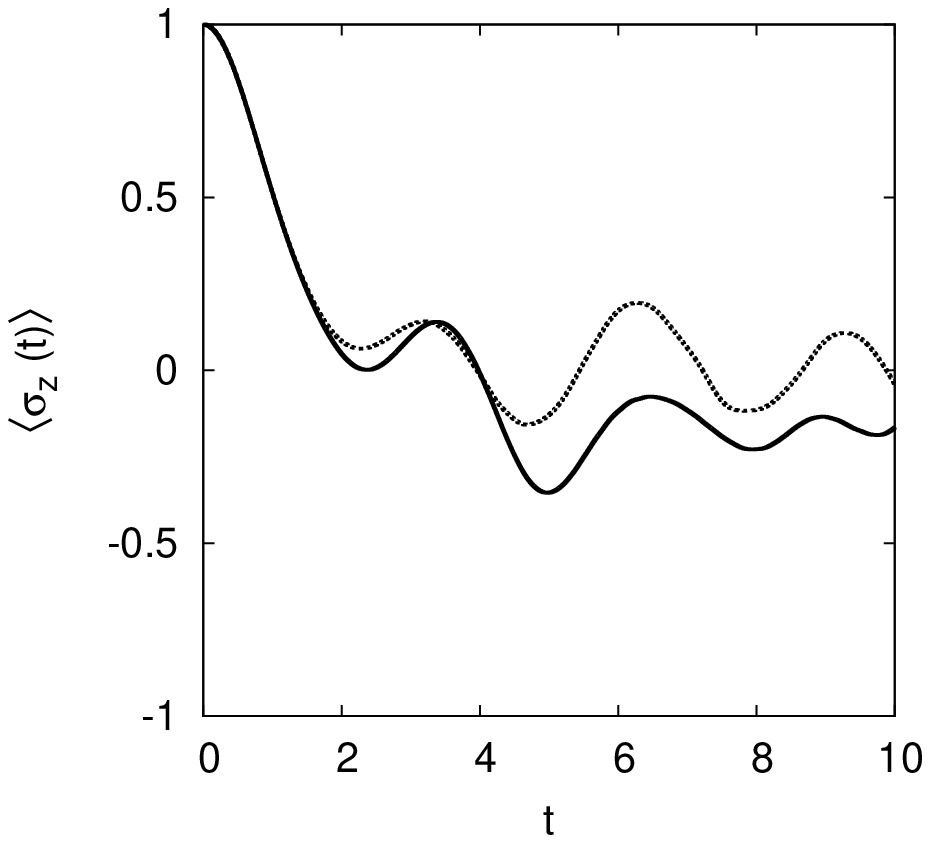}
\caption{Time dependence of the population $ \langle \sigma_z(t) \rangle$.
The dotted line corresponds to the quantum-classical evolution
while the solid-line corresponds to the quantum-corrected evolution.
Frequency ratio $\Omega = 0.6$, coupling constant $c = 0.6$,
initial average position $R_0 = -1.6$, harmonic coefficient $b_2 = -1$
and non-linearity parameter $b_4 = 0.5$.
}
\label{fig:fig4}
\end{figure}

\begin{figure}
\includegraphics[scale=0.7]{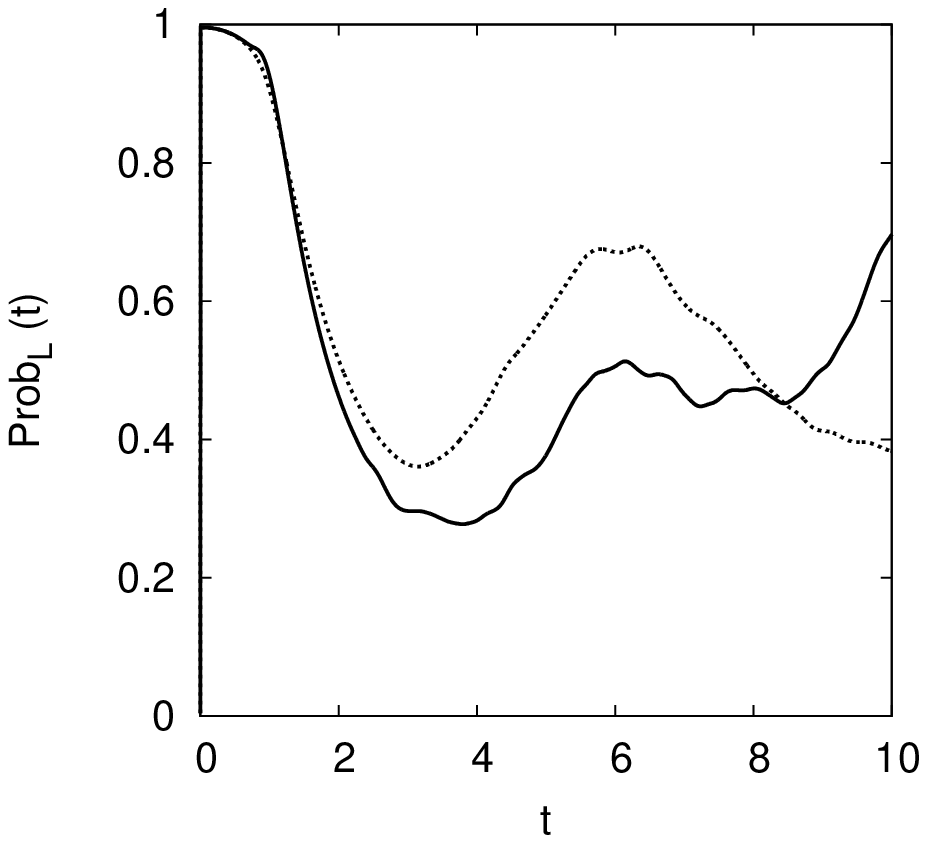}
\caption{Time dependence of instantaneous oscillator left-well 
occupation probability Prob$_L$($t$).
The dotted line corresponds to the quantum-classical evolution
while the solid-line corresponds to the quantum-corrected evolution.
Frequency ratio $\Omega = 0.6$, coupling constant $c = 0.6$,
initial average position $R_0 = -1.6$, harmonic coefficient $b_2 = -1$ 
and non-linearity parameter $b_4 = 0.5$}
\label{fig:fig5}
\end{figure}

\begin{figure}
\includegraphics[scale=0.7]{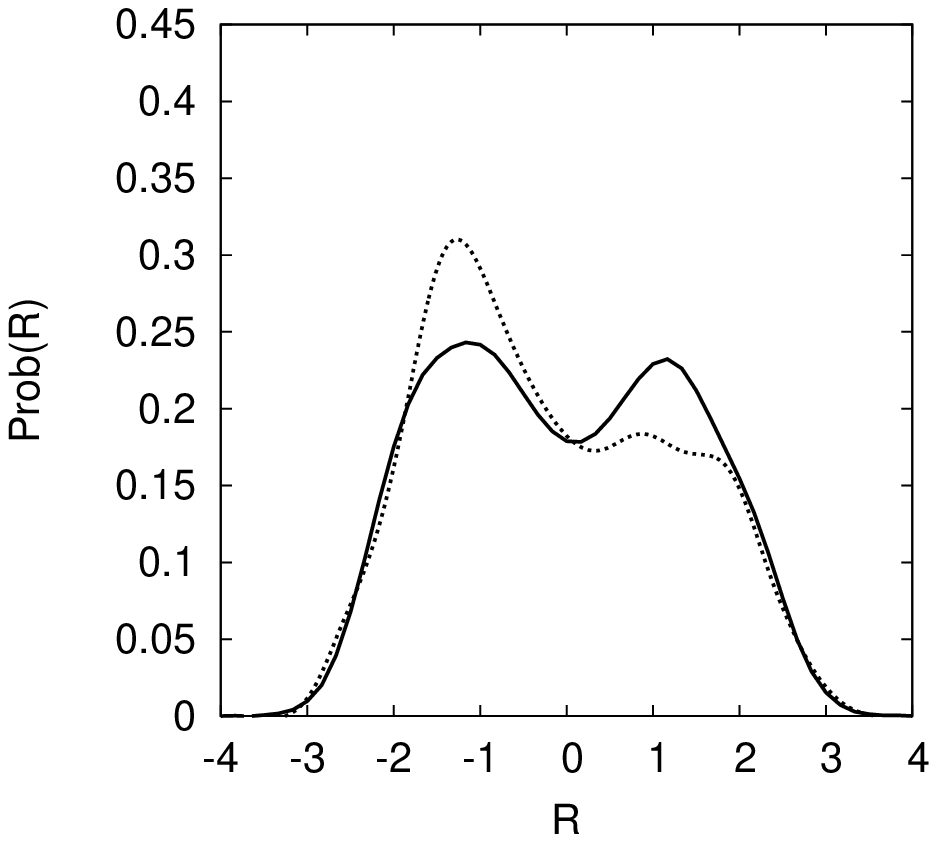}
\caption{Oscillator position-probability Prob($R$).
The dotted line corresponds to the quantum-classical evolution
while the solid-line corresponds to the quantum-corrected evolution.
Frequency ratio $\Omega = 0.6$, coupling constant $c = 0.6$,
initial average position $R_0 = -1.6$,
harmonic coefficient $b_2 = -1$ and non-linearity parameter $b_4 = 0.5$}
\label{fig:fig6}
\end{figure}

The increased damping effect is clear in Fig.~\ref{fig:fig1} for $t \geq 5$, 
showing the results obtained for $c = 0.4$. The pronounced damping near $t = 10$ 
corresponds to the highly well-symmetric phase-space configuration, in contrast to the
asymmetry of the classical case, 
intimated by the instantaneous left-well occupation probability
in Fig.~\ref{fig:fig2} at around $t \geq 7$.  Figure.~\ref{fig:fig3} shows the 
time-averaged position probability
of the quartic oscillator in the same calculation: the asymmetry in classical results, in contrast with
the symmetry of the quantum case, 
again demonstrates that the quantum corrections introduce 
additional tunnelling effects into the dynamics. 
Seemingly, tunnelling effects are not absent from the classical case. However, such effects
are the result of non-adiabatic interactions with the pseudo-spin, which are proportional in
magnitude to the strength of the coupling between systems.
Many of the general characteristics of such findings are confirmed
at higher couplings. 
The increased damping effects are still apparent in the results for $c= 0.6$, 
as seen in Fig.~\ref{fig:fig4} for $t \geq 5$. 
The instantaneous occupation probabilities in Fig.~\ref{fig:fig5} show clear
 differences between the quantum and the classical cases, 
larger fluctuations in the quantum result demonstrates
that greater tunnelling effects are still in evidence.
Additionally the pronounced damping occurring for $t \geq 5$
corresponds strongly to a period of highly symmetric well occupation in the quantum case. 
This calculation also displays a classical well-asymmetry 
in Fig.~\ref{fig:fig6} but of reduced magnitude when compared to that in Fig.~\ref{fig:fig3}.
%
%
%
%
%
%
%
%
%
% This can be seen somewhat in the time-averaged position-probability profiles, 
%comparing Figures.~(\ref{fig:fig3}) and (\ref{fig:fig6}),
%one can see that: as the coupling is increased, 
%the classical profile begins to converge with its quantum counterpart.
%
%
Comparison of Figs.~\ref{fig:fig1} and \ref{fig:fig4}, 
in conjunction with \ref{fig:fig2} and \ref{fig:fig5} demonstrate 
the importance of tunnelling effects in distinguishing quantum behaviour in the 
quartic oscillator through the population difference of the pseudo-spin. 
As the differences in the occupation probability decrease, so do the 
short-time qualitative differences in the behaviour of the Rabi 
oscillations. The instantaneous well-occupation probability plots also demonstrate
a correlation between differences in the width of the phase-space
distribution and differences in the damping effects on the Rabi oscillations.

\begin{figure}
\includegraphics[scale=0.7]{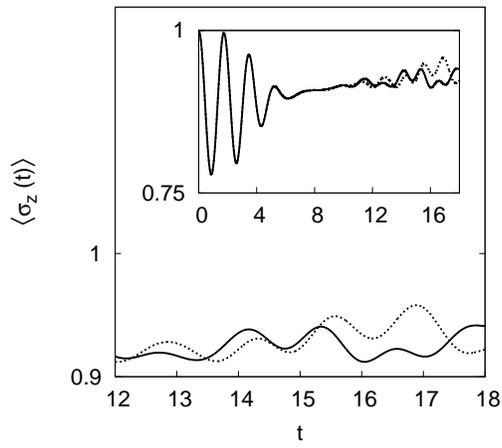}
\caption{Time dependence of the population $\langle \sigma_z(t) \rangle$. 
The dotted line corresponds to the quantum-classical evolution
while the solid-line corresponds to the quantum-corrected evolution. 
The inset displays the full time interval.
Frequency ratio $\Omega = 0.6$, coupling constant $c = 0.1$,
initial average position $R_0 = -7.0$, harmonic coefficient $b_2 = -0.01$
and non-linearity parameter $b_4 = 0.0004$.}
\label{fig:fig7}
\end{figure}

For $b_2=-0.01$ and $b_4=0.0004$, values which are in agreement with
the elastic theory predictions for the properties of the nano-rod,
we observe dynamical effects which are similar to those obtained
for the first set of values, although on a longer time scale. 
Figure~\ref{fig:fig7} displays the results of this calculation.
The variation of the population of the two-level system is somewhat smaller
than that achieved with the previous set of parameters.
The necessity of integrating the equations for longer times also puts
the numerical algorithm under strain and we find that the numerical
error increases too much beyond $t=18$. This is just a reflection
of the general difficulty in the integration of long time quantum dynamics.
Nevertheless, there are clear  indications that the damping of
the population is stronger in the case of a quantum motion
of the nano-rod.

%%%%%%%%%%%%%%%%%%%%%%%%%%%%%%%%%%%%%%%%%

\section{Conclusion}

This work suggests that
quantum effects in the motion of nano-rods under compression
can be detected indirectly.
The nano-rod can be coupled to a Cooper-pair box
and the tunnelling dynamics of the latter
can witness the quantum or classical features of
the time evolution of the first. In the case where the
nano-rod properties are far from elastic theory predictions, these 
differences emerge rapidly and greatly illuminate the relevant 
dynamics. However, one notes that it is
the non-linearity of the oscillations in the nano-rod
system that make this comparison of quantum and classical
dynamical effects possible, as this non-linearity
leads to the existence of quantum dynamical corrections. 
Despite this, the differences between quantum and classical evolutions persisted when the properties of the 
nano-rod where chosen in agreement with the elastic theory 
of materials, even though the non-linearity parameter is greatly 
reduced in this case. This is because the strain on the rod can be adjusted 
to ensure the non-linear effects are still detectable.
However, one must take into
account that this requires the nano-rod to be maintained around one hundredth above the critical strain
and that the dimensions of the nano-rod, necessarily having a length significantly greater than its thickness, may make
experimental control more difficult.

In the case where the nano-rod properties are in agreement with the predictions 
of elastic material theory, our calculations need to be performed for
a time scale larger by a factor of two and the variation of the two-level
system population is somewhat smaller. The longer time scale puts the calculation
in conflict with the general problem of the integration of long time quantum dynamics.
However, within the time span which is accessible in a reliable way to the
algorithm that we have used, the signs that allows one to discriminate between
the quantum and the classical motion of the nano-rod remain apparent.

In the future we plan to study alternative numerical schemes to extend 
the time span of the calculations even further as well as to study other 
schemes of detection based on multi-level systems.

\section{Acknowledgments}

This work is based upon research supported by
the National Research Foundation of South Africa.
Part of this work has been performed during a sabbatical stay
of A.S. at the Department of Physics of the University
of Messina in Italy.

%% References

%%%%%
%%%%%
%%%%%
%%%%%
%%%%%
%%%%%%
\end{document}